\documentclass[fleqn,12pt,twoside]{article}
\usepackage{espcrc1}

\usepackage{graphicx}
\usepackage[figuresright]{rotating}

\newcommand{\AmS}{{\protect\the\textfont2
  A\kern-.1667em\lower.5ex\hbox{M}\kern-.125emS}}

\title{Centrality dependence of the multiplicity and transverse momentum distributions
at RHIC and LHC and the percolation of strings}

\author{M. A. Braun\address{Department of High Energy Physics, University
of St. Petersburg, \\
        19894 St. Petersburg, Russia},
        F. del Moral\address{Departamento de F\'{\i}sica de Part\'{\i}culas
        Elementales, Univ. Santiago de Compostela, \\
        15706 Santiago de Compostela, Spain},
        {\underline{C. Pajares}}\addressmark}

\begin{document}

% typeset front matter
\maketitle

\begin{abstract}
The dependence of the multiplicity and the transverse momentum
distribution on the number of collisions are studied for central
and peripheral Au-Au collisions at SPS, RHIC and LHC energies in
the framework of percolation of strings. A scaling law relating
the multiplicity to the mean transverse momentum is obtained. Our
results are in overall agreement with the SPS and RHIC data,
obtaining a suppression on $p_T$ distribution even for $p_T$
larger than 1 GeV/c.
\end{abstract}

\section{INTRODUCTION}
Most of the multiparticle production is currently described in
terms of color strings stretched between the projectile and
target, which decay into new strings and subsequently hadronize to
produce observed hadrons. Color strings may be viewed as small
areas in the transverse space, $\pi r_o^2$, $r_o$=0.2 fm, filled
with color field created by the colliding partons. Particles are
produced via emission of $q\bar q$ pairs in this field
\cite{cita1}. With growing energy and/or atomic number, the number
of strings grows, and they start to overlap, forming clusters,
very much similar to disks in the two dimensional percolation
theory \cite{cita2}. At a certain critical density,
\begin{equation}\eta_c=N_s \pi r_o^2/S\label{ec1}\end{equation}
a macroscopic cluster appears that marks the percolation phase
transition. In (\ref{ec1}) $N_s$ is the number of strings and $S$
is the total transverse area of the scattering of the colliding
nuclei.

The percolation theory governs the geometrical pattern of the
string clustering. Its observable implications, however, require
introduction of some dynamics to describe string interaction,
i.e., the behavior of a cluster formed by several overlapping
strings. There are several possibilities \cite{cita3,cita4}. Here,
we assume that a cluster behaves as a single string with a higher
color field $\vec Q_n$ corresponding to the vectorial sum of the
color charge of each individual $\vec Q_1$ string. The resulting
color field covers the area $S_n$ of the cluster. As $\vec
Q_n=\sum \vec Q_1$, and the individual string colors may be
oriented in an arbitrary manner respective to one another, the
average $\vec Q_{1i}\vec Q_{1j}$ is zero and $\vec Q_n^2=n\vec
Q_1^2$.

Knowing this charge color $\vec Q_n$, one can compute the particle
spectra produced by a single color string of area $S_n$ using the
Schwinger formula \cite{cita5}. For the multiplicity $\mu_n$ and
the average $p_T^2$ of particles $<p_T^2>_n$ produced by a cluster
of $n$ strings one finds
\begin{equation}\mu_n=\sqrt{nS_n\over S_1}\mu_1 \quad ; \quad
<p_T^2>_n=\sqrt{nS_1\over S_n} <p_T^2>_1\label{ec2}\end{equation}
where $\mu_1$ and $<p_T^2>_1$ are the mean multiplicity and
$p_T^2$ of particles produced by a simple string with area
$S_1=\pi r_o^2$. For strings, just touching each other $S_n=nS_1$
and hence $\mu_n=n\mu_1$, $<p_T^2>_n=<p_T^2>_1$, as expected. In
the opposite case of maximum overlapping $S_n=S_1$ and therefore
$\mu_n=\sqrt{n}\mu_1$, $<p_T^2>=\sqrt{n}<p_T^2>_1$

Equation (\ref{ec2}) is the main tool of our calculation
\cite{cita6}. In order to compute the multiplicities we generate
strings according the Monte Carlo code of \cite{cita1}. Each
string is produced at an identified impact parameter. From this,
knowing the transverse area of each string, we identify all the
clusters formed in each collision and subsequently compute for
each of them its multiplicity in units $\mu_1$. The value of
$\mu_1$ was fixed by normalizing our results to the SPS WA98
results for central Pb-Pb collisions \cite{cita7}. In color
strings models both $\mu_1$ and $<p_T^2>$ are assumed to rise with
energy due to increase of the rapidity interval available and to
the hard contribution. However this rise is very weak in the range
$\sqrt{s}$=17-200 GeV and we can neglect it, in the first
approximation.

\section{RESULTS}
\begin{figure}[!h]
 \begin{center}
    \includegraphics[scale=0.58]{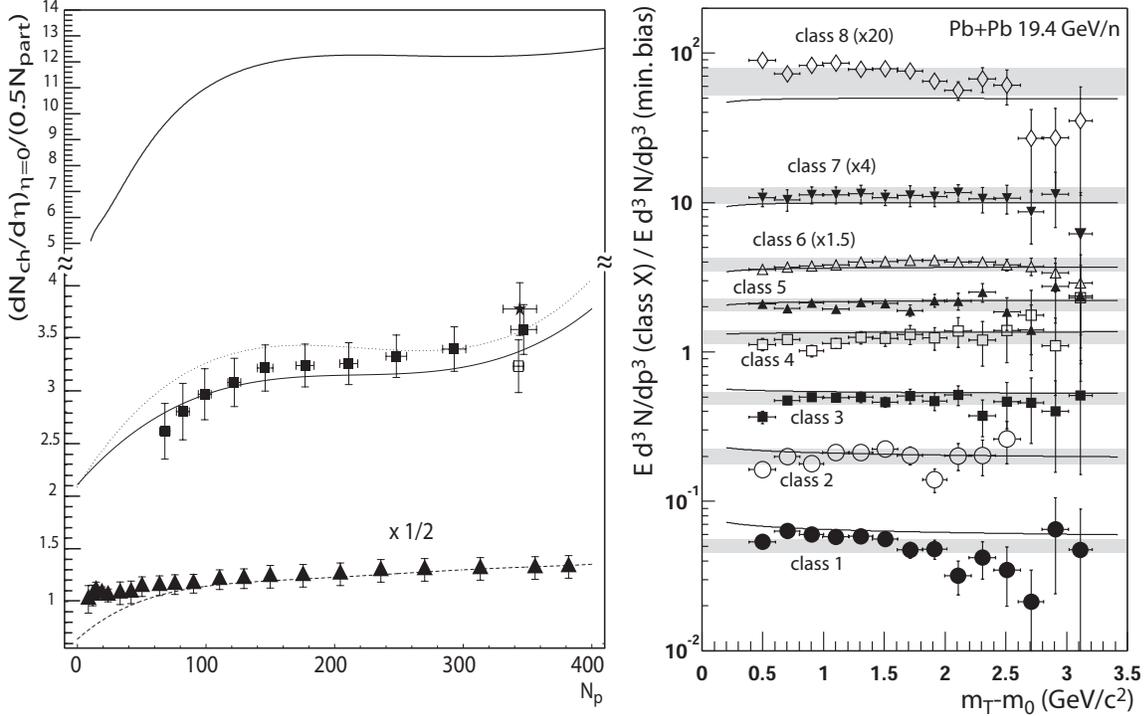}
    \caption{ \small a) Comparison of our predictions of the multiplicity with the
    number of participants, the SPS WA98 \cite{cita7} (filled triangles),
    the RHIC Phenix \cite{cita8} (filled boxes), Phobos \cite{cita9}
    (nonfilled box) data at $\sqrt{s}$=130 GeV/n, and RHIC Phenix data
    at $\sqrt{s}$=200 GeV/n (filled star). The upper solid line represents our
    prediction for Pb-Pb collisions at 5500 GeV/n. b) Comparison of our predictions of
    ratios of invariant multiplicities of neutral pions for Pb-Pb collisions of
    different centralities to minimum bias distributions as a function of $m_T-m_o$
    and SPS WA98 data. Meaning of centrality classes and grey bands can be found
    en reference \cite{cita11}.  }
    \label{figure1}
 \end{center}
\end{figure}

The comparison of our results for the dependence of the
multiplicity on the number of participants with the SPS WA98 data,
the RHIC Phenix \cite{cita8} and Phobos \cite{cita9} data at
$\sqrt{s}$=130 GeV and $\sqrt{s}$=200 GeV, is presented in Fig.
\ref{figure1}.a, where also is presented our prediction for LHC.

In order to compute the transverse momentum distribution we make
use of the parametrization of the $pp$ data at 130 GeV
\begin{equation}{dN\over dp_T^2}={a\over (p_o
+p_T)^\alpha}\label{ec3}\end{equation} where $a$,$p_o$ and
$\alpha$ are parameters fitted to data. The standardly expected
distribution for Au-Au collision is
\begin{equation}{dN\over dp_T^2}=<N_{coll}>{dN\over dp_T^2}\Big
|_{pp}\label{ec4}\end{equation}

However, from Eq. (\ref{ec1}) we have
\begin{equation}<p_T^2>_{Au-Au}=<p_T^2>_{pp}{<nS_1/S_n>_{Au-Au}^{1/2}\over
<nS_1/S_n>_{pp}^{1/2}}\end{equation}

\begin{figure}[!h]
 \begin{center}
    \includegraphics[scale=0.85]{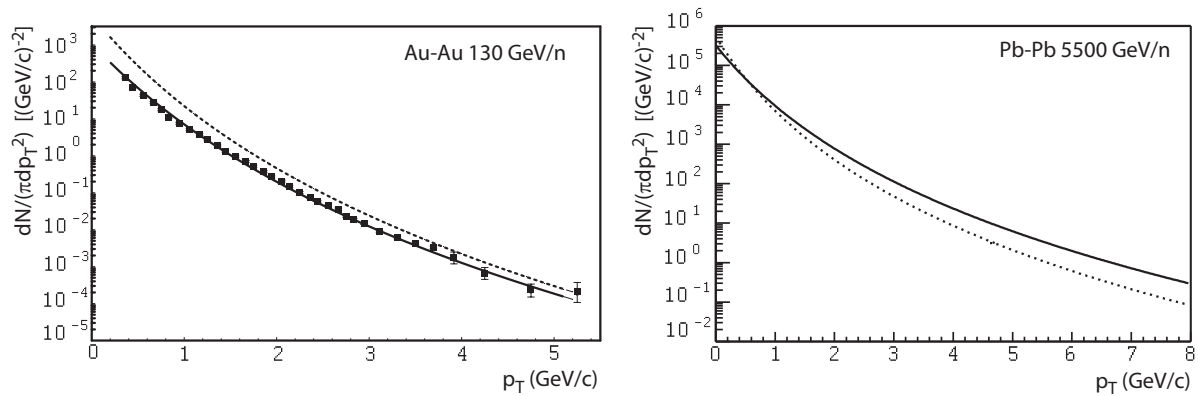}
    \caption{ \small Expected $p_T$ distribution using the percolation of strings
    model (solid line) for 5\% central Au-Au and Pb-Pb collisions at 130 GeV/n
    and 5500 GeV/n respectively. Also it is shown the expected distribution from
    the independent string picture (dotted line).}
    \label{figure2}
 \end{center}
\end{figure}

This implies that the same parametrization (\ref{ec3}) can be used
for nucleus-nucleus collisions with the only change
\begin{equation}p_o\rightarrow p_o\Big ( {<nS_1/S_n>_{Au-Au}\over
<nS_1/S_n>_{pp}}\Big )^{1/4}\label{ec6}\end{equation} In Fig.
\ref{figure1}.b we show the comparison of our results for eight
different centrality events in Pb-Pb collisions with WA98 data
\cite{cita11}. In Fig. \ref{figure2}.a we show the distribution
(\ref{ec3}) for charged particles for central (5\%) Au-Au
collisions compared to the Phenix data \cite{cita10}. Also the
distribution expected from the independent string picture, Eq.
(\ref{ec4}), is shown. A very good agreement is obtained. We also
obtain a good agreement for peripheral collisions, see reference
\cite{cita6}. In Fig. \ref{figure2}.b we show our prediction for
central Au-Au collisions at $\sqrt{s}$=5500 GeV compared with the
formula (\ref{ec4}). In this case, contrary to RHIC energy, a
crossing at low $p_T$ is predicted.

We observe that for a cluster, $<p_T^2>={S_1\over
S_n}{<p_T^2>_1\over \mu_1}\mu_n$, which at high density translate
into
\begin{equation}<p_T^2>_{AA}={\pi r_o^2\over S_{AA}}
{<p_T^2>_1\over \mu_1} \mu_{AA}\label{ec7}\end{equation} The
formula (\ref{ec7}) shows an universal relation between
multiplicities and $<p_T^2>$. The scaling laws expressed by the
formula (\ref{ec3}) with the change (\ref{ec6}) and by the formula
(\ref{ec7}) are very similar to other scaling laws using
saturation of gluons.

To conclude, a general comment. We obtain a good description of
the multiplicities and transverse momentum distribution at
different energies and different centralities including a low and
high $p_T$ suppression. However we can not reproduce the
additional high $p_T$ suppression reported at this conference by
several collaborations. If the explanation is jet quenching, our
approach can provide an unified framework to add the phenomenology
of jet quenching. This work was done under contract FPA2002-01161
of Ministerio de Ciencia y Tecnologia of Spain.

\end{document}